\documentclass[conference]{IEEEtran}
\IEEEoverridecommandlockouts
\usepackage{cite}
\usepackage{amsmath,amssymb,amsfonts}
\usepackage{algorithmic}
\usepackage{graphicx}
\usepackage{textcomp}
\usepackage{xcolor}
\usepackage{tcolorbox}
\usepackage{subcaption}
\usepackage{enumerate}
\usepackage{enumitem}
\usepackage{hyperref}
\usepackage{gensymb}
\def\BibTeX{{\rm B\kern-.05em{\sc i\kern-.025em b}\kern-.08em
    T\kern-.1667em\lower.7ex\hbox{E}\kern-.125emX}}

\begin{document}

\title{Potential and Limitation of High-Frequency Cores and Caches
}

\author{\IEEEauthorblockN{Kunal Pai}
\IEEEauthorblockA{
\textit{University of California, Davis}\\
Davis, USA \\
kunpai@ucdavis.edu}
\and
\IEEEauthorblockN{Anusheel Nand}
\IEEEauthorblockA{
\textit{University of California, Davis}\\
Davis, USA \\
anunand@ucdavis.edu}
\and
\IEEEauthorblockN{Jason Lowe-Power}
\IEEEauthorblockA{
\textit{University of California, Davis}\\
Davis, USA \\
jlowepower@ucdavis.edu}
}
% \and
% \and
% \and

\maketitle

\begin{abstract}
This paper explores the potential of cryogenic semiconductor computing and superconductor electronics as promising alternatives to traditional semiconductor devices.
As semiconductor devices face challenges such as increased leakage currents and reduced performance at higher temperatures, these novel technologies offer high performance and low power computation.
Conventional semiconductor electronics operating at cryogenic temperatures (below -150\degree~C or 123.15~K) can benefit from reduced leakage currents and improved electron mobility.
%Cryogenic computing operates at ultra-low temperatures near 77~K, leading to lower leakage currents and improved electron mobility.
On the other hand, superconductor electronics, operating below 10~K, allow electrons to flow without resistance, offering the potential for ultra-low-power, high-speed computation.
This study presents a comprehensive performance modeling and analysis of these technologies and provides insights into their potential benefits and limitations.
We implement models of in-order and out-of-order cores operating at high clock frequencies associated with superconductor electronics and cryogenic semiconductor computing in gem5.
We evaluate the performance of these components using workloads representative of real-world applications like NPB, SPEC CPU2006, and GAPBS.
Our results show the potential speedups achievable by these components and the limitations posed by cache bandwidth.
This work provides valuable insights into the performance implications and design trade-offs associated with cryogenic and superconductor technologies, laying the foundation for future research in this field using gem5.
Our code and data are open-source and available on GitHub\footnote{\href{https://github.com/darchr/gem5-cryo-superconducting}{https://github.com/darchr/gem5-cryo-superconducting}}.
\end{abstract}

% \begin{IEEEkeywords}
% Cryogenic Computing, Superconducting Electronics, gem5, Performance Modeling, Workload Analysis
% \end{IEEEkeywords}

\section{Introduction}

Traditional semiconductor devices suffer from increased leakage currents and reduced performance as temperatures rise~\cite{cryocore}, leading to significant energy dissipation and limiting the scalability of modern computing systems.
Coupling this with a slowing Moore's Law~\cite{theis2017end}, the performance improvements of traditional computing systems have stagnated.
As a countermeasure, researchers are exploring novel computing technologies which promise high performance and low power computation.

Conventional semiconductor electronics operating at cryogenic temperatures (below -150\degree~C or 123.15~K), present a promising avenue on high-speed and low-power computation~\cite{cryocore}.
% At cryogenic temperatures, semiconductor devices exhibit lower leakage currents and improved electron mobility, leading to improved performance~\cite{cryocore}.
Another technology, superconductor electronics, operating below 10~K, offers the potential for ultra-low-power, high-speed computation, by allowing electrons to flow without resistance~\cite{devoret2013superconducting}.
% As a countermeasure, researchers have explored the potential of cryogenic computing, which operates at ultra-low temperatures near 77 K, to mitigate these issues and unlock new opportunities for high-performance computing.
%Furthermore, operating at 0 K, superconducting electronics offer the potential for no resistance and ultra-low-power, high-speed computation.
% Therefore, both these technologies have the potential to provide unprecedented speedups to computation.
These technologies allow for individual components to operate at a higher clock frequency than typical CMOS devices.
% For certain workloads, this can result in a substantial speedup, as the components are able to complete their tasks faster.

% However, these performance improvement can be limited by the other components in the system.
% One example is if the other components in the system are not able to keep up with the increased clock frequency of the core, meaning that every time the core needs to access these components, it has to wait for them to respond.
% Therefore, the number of cycles required to complete the workloads increases, thereby limiting the performance improvement.
% Another example is if the cache bandwidth is not high enough.
% In this case, the core would not be capable of achieving the expected number of cache accesses within unit time due to the increased clock frequency, thereby limiting the performance improvement in real-world scenarios.
% Another example is if the workload is memory-intensive.
% Since the memory operates at room temperature, any memory-intensive workload would be bottlenecked by the memory, as the core would have to wait for the memory to respond to its requests, implying that the workload would not be able to achieve the speedup that the faster core could achieve.
% Therefore, speedups of workloads are closely related to the bandwidths of the caches and the other components in the system.

% While there is prior work investigating architecture and implementation of components in cryogenic and superconducting technologies, to the best of our knowledge, there is no prior work looking at 
In this paper, we study the full-system implications of running the processing core and caches at ultra-high frequencies.
Specifically, we introduce a modeling framework and perform and initial evaluation of out-of-order and in-order cores at high clock frequencies associated with superconductor electronics and cryogenic semiconductor computing~\cite{IEEE2023Cryogenic}. %, and give us the necessary values to calculate if the maximum possible speedup for a workload is truly achievable.
We also wish to substantiate how other components in the system affect the performance, and how these other components should be designed to keep up with the faster components.

% \note{Add a paragraph on implementation, gem5, and workloads.}

To evaluate the performance of cryogenic semiconductor and superconductor computing technologies, we:
\begin{enumerate}[leftmargin=*]
    \item Use gem5~\cite{lowe2020gem5,binkert2011gem5} to implement models of in-order and out-of-order cores operating at high clock frequencies, typical of cryogenic semiconductor and superconductor environments.
    \item Choose workloads representative of real-world applications like NPB~\cite{npb}, SPEC CPU2006~\cite{henning2006spec}, and GAPBS~\cite{beamer2015gap} to evaluate the performance of these components.
    These workloads provide a comprehensive understanding of how different types of applications respond to the ultra-fast processing capabilities offered by cryogenic semiconductor and superconductor technologies.
    \item Simulate these workloads in different combinations of the aforementioned components in a superconductor clock domain and in the cryogenic semiconductor clock domain, while keeping the board and memory at their room temperature frequencies.
    \item Show the potential speedups we can get from the combinations along with the potential limitation posed by the cache bandwidth required for that speedup.
\end{enumerate}

This is a preliminary study to understand the potential of using superconductor electronics for general-purpose computing.
Our results indicate that superconductor electronics may not be well suited for general-purpose computing, as the cache bandwidth limits the performance.
However, we believe that superconductor electronics can be used for some applications.
We conclude future work should focus on designing specialized accelerators that can take advantage of the potential speedups offered by superconductor electronics, instead of using them for general-purpose computing.

% To evaluate the performance of cryogenic and superconducting computing technologies, we use gem5~\cite{lowe2020gem5,binkert2011gem5}, a widely-used simulator for computer architecture research.
% Using gem5, we implement models of both in-order and out-of-order cores operating at the high clock frequencies typical of cryogenic and superconducting environments.
% We choose workloads representative of real-world applications like NPB~\cite{npb}, SPEC CPU2006~\cite{henning2006spec}, and GAPBS~\cite{beamer2015gap} to evaluate the performance of these components.
% These workloads provide a comprehensive understanding of how different types of applications respond to the ultra-fast processing capabilities offered by cryogenic and superconducting technologies.
% We then simulate these workloads in different combinations of the aforementioned components in a superconducting clock domain and in the cryogenic clock domain, while keeping the board and memory at their room temperature frequencies.
% We then show the potential speedups we can get from the combinations along with the potential limitation posed by the cache bandwidth required for that speedup.

\section{Background}

In recent years, the pursuit of novel computing paradigms has led to a resurgence of interest in cryogenic semiconductor components and superconductor circuits.
These technologies promise unprecedented speedups by exploiting the unique properties of materials at extremely low temperatures.

While superconducting was first discovered in 1911~\cite{onnes1911resistance}, the theoretical concepts behind it were not fully understood until the BCS theory of the 1950s~\cite{bardeen1957theory}.
Due to further work in concepts like the Josephson effect~\cite{josephson1962possible} and flux quantization~\cite{doll1961experimental}, superconductor circuits have been understood to operate at high clock frequencies of up to 100~GHz~\cite{chen1999rapid}.
Therefore, superconductor circuits enable ultra-fast signal processing with minimal energy dissipation \cite{devoret2013superconducting}.
% Due to effects of flux quantization~\cite{doll1961experimental} and the Josephson effect~\cite{josephson1962possible}, superconducting circuits can operate at clock frequencies of up to 100 GHz~\cite{chen1999rapid}.

Cryogenic semiconductor components offer improved performance and energy efficiency by the reduced resistance at ultra-low temperatures.
At cryogenic temperatures, semiconductor devices exhibit lower leakage currents and improved electron mobility, leading to improved performance.
Therefore, these components also operate at a higher clock frequency, usually around 4~GHz~\cite{cryocore}.

Cryogenic semiconductor computing has gained significant attention in recent years, with researchers exploring cryogenic semiconductor components to enhance computing performance.
Byun et al. introduced CryoCore, a cryogenic core based on the BOOM core, operating at 4.0~GHz with improved performance at cryogenic temperatures~\cite{cryocore}.
Min et al. proposed CryoCache, a cryogenic cache hierarchy leveraging these temperatures to enhance cache performance~\cite{min2020cryocache}.
These components operate at ultra-low temperatures and offer substantial speedups compared to conventional semiconductor devices.

Our work represents a crucial step towards the practical implementation of superconductor and cryogenic semiconductor components in computing systems. Notably, we:
\begin{enumerate}[leftmargin=*]
    \item Demonstrate the feasibility of integrating high-level characteristics of these components into gem5~\cite{lowe2020gem5,binkert2011gem5}.
    \item Show the potential speedups and limitations of high-frequency components, providing valuable insights into the performance implications and design trade-offs associated with these novel technologies.
    \item Lay the foundation for gem5 as a modeling framework for this field of computing research, providing all the tools necessary for evaluating the performance of computing systems under cryogenic and superconductor conditions.
\end{enumerate}

\section{Methodology}

For all the experiments, we used gem5~\cite{binkert2011gem5,lowe2020gem5} as the simulator.
% gem5 is a cycle-level simulator that provides numerous CPU models, cache models, and memory models.
We leverage the easy modification of gem5 CPU and cache models to make our own custom components, which is essential due to the microarchitectural details of the cryogenic semiconductor computing models available in the literature.
We can also place the CPU, caches, and memory in different clock domains, which is essential for simulating cryogenic semiconductor computing environments.
The gem5 Resources artifacts~\cite{bruce2021gem5art} provides workloads that we can also use for our experiments.
Hence, gem5 is the ideal choice for our experiments.

% We assume that the only modification that gem5 requires to run at cryogenic and superconducting environments is to change the clock frequency.
% Therefore, we are not considering any power models or area models in our simulations.
% \note{I'm not sure what this paragraph is trying to say.}

\subsection{Core Microarchitecture}

We modeled the cryogenic semiconductor core and caches by building a model based on CryoCore by Byun et al.~\cite{cryocore}.
Their cryogenic core was based on BOOM~\cite{celio2015berkeley}, an out-of-order RISC-V core, and could run at 4.0 GHz.
We used microarchitectural details provided in their paper to create a variant of the O3CPU in gem5.
These microarchitectural details are enumerated in Table~\ref{tab:gem5_parameters_cryocore}.
For the pipeline values that were not specified, we went with the default values provided by gem5 for those variables.
With regard to the latencies of the functional units and the branch predictor, we decided to go with the ones provided as part of the \textit{RISCVMatched} prebuilt board~\cite{riscvmatched} (based on the HiFive Unmatched and having a similar architecture to Rocket and BOOM~\cite{asanovic2016rocket}), since these values have been partially validated~\cite{kunal2023matchedposter}.

We also created an in-order variant of the CryoCore, called the In-Order CryoCore, which is based on the pre-built \textit{RISCVMatched} board~\cite{riscvmatched} in gem5.
This core is based on the HiFive Unmatched core, which is an in-order core.
The pipeline comprises of eight stages: two stages of instruction fetch (F1 and F2), two stages of instruction decode (D1 and D2), address generation (AG), two stages of data memory access (M1 and M2), and register write-back (WB)~\cite{siFiveFU740Manual}.
Execution takes place either in the AG stage or the M2 stage, depending on the instruction.
This core has been partially validated~\cite{kunal2023matchedposter}.

The latencies of the functional units and the branch predictor for both the CryoCore and the In-Order CryoCore are the same as the ones provided in the \textit{RISCVMatched} board~\cite{riscvmatched}.
These values can be viewed in Table~\ref{tab:gem5_parameters_common}.
% The microarchitectural details of this core can be viewed in Table ~\ref{tab:gem5_parameters_cryocore_inorder}.

\begin{table}[h]
    \centering
    \caption{Microarchitectural details of the gem5 model of CryoCore\protect\footnotemark}
    \label{tab:gem5_parameters_cryocore}
    \begin{tabular}{|c|c|}
        \hline
        \textbf{Parameter} & \textbf{Value in gem5 Model} \\
        \hline
        Cache Load/Store Ports & 1 \\
        \hline
        Instruction Width & 4 bytes \\
        \hline
        Fetch Queue Size & 24 \\
        \hline
        Load/Store Queue Entries & 24 \\
        \hline
        Instruction Queue Entries & 72 \\
        \hline
        Reorder Buffer Entries & 96 \\
        \hline
        Integer Registers & 180 \\
        \hline
        Floating Point Registers & 168 \\
        \hline
    \end{tabular}
\end{table}
\footnotetext{Found in \href{https://github.com/darchr/gem5-cryo-superconducting/blob/main/components/cryocore/cryocore.py}{components/cryocore/cryocore.py}}.

\begin{table}[h]
    \centering
    \caption{Common microarchitectural details of the gem5 model of CryoCore and In-Order CryoCore\protect\footnotemark}
    \label{tab:gem5_parameters_common}
    \begin{tabular}{|c|c|}
        \hline
        \textbf{Parameter} & \textbf{Value in gem5 Model} \\
        \hline
        BTB Entries & 32 \\
        \hline
        RAS Entries & 12 \\
        \hline
        Branch Predictor Size & 16 KB \\
        \hline
        History Table Size & 4 KB \\
        \hline
        Indirect Branch Predictor Size & 16 entries \\
        \hline
        Branch Predictor Counter Bits & 4 \\
        \hline
        Integer FU Latency & 1 \\
        \hline
        Multiplication FU Latency & 3 \\
        \hline
        Division FU Latency & 6 \\
        \hline
        Memory Read/Write Latency & 2 \\
        \hline
    \end{tabular}
\end{table}
\footnotetext{Found in \href{https://github.com/darchr/gem5-cryo-superconducting/blob/main/components/cryocore/cryocore.py}{components/cryocore/cryocore.py}}.

\subsection{Cache Microarchitecture}

We modelled the cache hierarchy based on CryoCache by Min et al.~\cite{min2020cryocache}.
Their design has a private L1 cache, a private L2 cache, and a shared L3 cache.
We used microarchitectural details provided in their paper to create a variant of the cache hierarchy in gem5 with those details.
They can be viewed in greater detail in Table~\ref{tab:gem5_parameters_cryocache}.

\begin{table}[h]
    \centering
    \caption{Microarchitectural details of the gem5 model of CryoCache\protect\footnotemark}
    \label{tab:gem5_parameters_cryocache}
    \begin{tabular}{|c|c|}
        \hline
        \textbf{Parameter} & \textbf{Value in gem5 Model} \\
        \hline
        L1D Cache Size & 32 kB \\
        \hline
        L1D Cache Associativity & 8 \\
        \hline
        L1D Cache Data Latency & 2 \\
        \hline
        L1I Cache Size & 32 kB \\
        \hline
        L1I Cache Associativity & 8 \\
        \hline
        L1I Cache Data Latency & 2 \\
        \hline
        L2 Cache Size & 512 kB \\
        \hline
        L2 Cache Associativity & 8 \\
        \hline
        L2 Cache Data Latency & 8 \\
        \hline
        L3 Cache Size & 16 MB \\
        \hline
        L3 Cache Associativity & 16 \\
        \hline
        L3 Cache Data Latency & 21 \\
        \hline
    \end{tabular}
\end{table}
\footnotetext{Found in \href{https://github.com/darchr/gem5-cryo-superconducting/blob/main/components/cryocache/cryocache.py}{components/cryocache/cryocache.py}}.

\subsection{Workloads}

For the workloads, we used a combination of small and large-sized benchmarks that are representative of real-world applications.
The small-sized workloads are provided as part of the ``\texttt{riscv-getting-started-benchmark-suite}''~\cite{riscv-getting-started-benchmark-suite} suite provided as part of gem5 Resources.
This suite's constituent workloads have been cherry-picked from popular benchmarks and applications.
These workloads are as follows:
\begin{itemize}
    \item \texttt{bfs} - Breadth First Search from the GAP benchmark suite ~\cite{beamer2015gap}.
    We used a graph with 1024 vertices and 10 iterations.
    \item \texttt{tc} - Triangle Counting from the GAP benchmark suite ~\cite{beamer2015gap}.
    We used a graph with 1024 vertices and 10 iterations.
    \item \texttt{minisat} - A SAT solver from the LLVM test suite ~\cite{llvm}.
    We used a SAT problem of 15000 variables and 20000 clauses.
    \item \texttt{is} - Integer Sort from the NAS Parallel Benchmarks ~\cite{npb}.
    We used the class S version of this workload.
    \item \texttt{lu} - Lower-Upper Gauss-Seidel from the NAS Parallel Benchmarks ~\cite{npb}.
    We used the class S version of this workload.
    \item \texttt{cg} - Conjugate Gradient from the NAS Parallel Benchmarks ~\cite{npb}.
    We used the class S version of this workload.
    \item \texttt{bt} - Block Tri-Diagonal from the NAS Parallel Benchmarks ~\cite{npb}.
    We used the class S version of this workload.
    \item \texttt{ft} - Fourier Transform from the NAS Parallel Benchmarks ~\cite{npb}.
    We used the class S version of this workload.
\end{itemize}

For the large workloads, we used a subset of the SPEC CPU2006~\cite{henning2006spec} benchmark suite, that is compatible with the RISC-V ISA in gem5.
These workloads are:
\begin{itemize}
    \item \texttt{400.perlbench} - A Perl interpreter.
    \item \texttt{401.bzip2} - A file compression utility.
    \item \texttt{410.bwaves} - A fluid dynamics simulation.
    \item \texttt{429.mcf} - A vehicle scheduling problem.
    \item \texttt{433.milc} - A lattice quantum chromodynamics simulation.
    \item \texttt{434.zeusmp} - A computational fluid dynamics simulation.
    \item \texttt{435.gromacs} - A molecular dynamics simulation.
    % \item \texttt{436.cactusADM} - A general relativity simulation.
    \item \texttt{437.leslie3d} - A fluid dynamics simulation.
    \item \texttt{444.namd} - A molecular dynamics simulation.
    \item \texttt{445.gobmk} - A game of Go AI.
    % \item \texttt{453.povray} - A ray tracing program.
    % \item \texttt{454.calculix} - A finite element analysis program.
    \item \texttt{456.hmmer} - A bioinformatics program.
    \item \texttt{458.sjeng} - A chess program.
    \item \texttt{459.GemsFDTD} - A finite-difference time-domain method simulation.
    \item \texttt{462.libquantum} - A quantum computer simulator.
    \item \texttt{464.h264ref} - An H.264 video encoder.
    % \item \texttt{465.tonto} - A quantum computer simulator.
    \item \texttt{470.lbm} - A lattice Boltzmann method fluid dynamics simulation.
    \item \texttt{471.omnetpp} - A discrete event network simulator.
    \item \texttt{473.astar} - A pathfinding program.
    % \item \texttt{482.sphinx3} - A speech recognition program.
\end{itemize}

The large workloads are computationally expensive, with complete executions taking months in gem5.
Therefore, to reduce simulation time, we used the SimPoints technique~\cite{sherwood2002automatically} to identify representative regions of the workloads, and used their weights to compute how the entire workload would perform.

\subsection{Experimental Setup}

\begin{figure}[h]
\centering
\includegraphics[width=\linewidth]{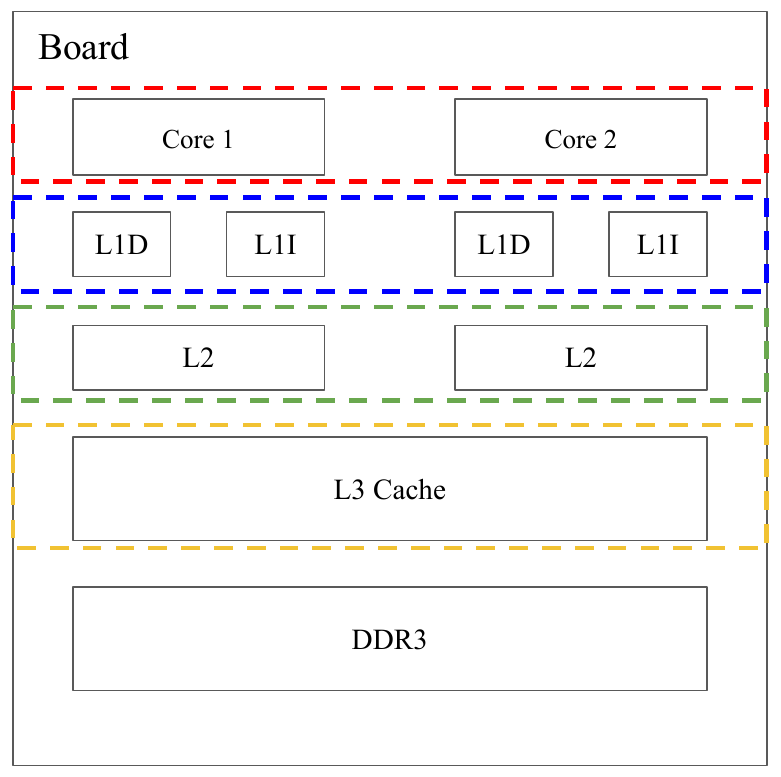}
\caption{Schematic representation of different clock domains of the board. Different colored rectangles represent different clock domains.}
\label{fig:clock}
\end{figure}

The experimental setup utilized a RISC-V system, or a ``board'', equipped with two CryoCores, each paired with a CryoCache.
The memory system was comprised of a single-channel DDR3\_1600\_8x8-based DIMM.

We categorized the board's subcomponents into distinct clock domains, as illustrated in Figure~\ref{fig:clock}.
Specifically, the CryoCore, L1 Caches, L2 Caches, and L3 Cache are segregated into different clock domains.
However, in all the experiments, the board itself operated at a constant room temperature clock frequency of 2 GHz, while the memory system maintained its own constant frequency of 800 MHz.

We conducted a series of experiments to evaluate the performance of the workloads under varying clock frequencies.
We varied the clock frequency of the CryoCore and the CryoCache in different combinations to observe the impact on the performance of the workloads.
These configurations are as follows:
\begin{itemize}
    \item \textit{CryoCore and CryoCache (CryoAll)} - Out-of-order CryoCore and CryoCache models in the cryogenic semiconductor clock domain (4~GHz).
    \item \textit{SuperCore and CryoCache (SuperCryo)} - CryoCore model in the superconductor clock domain (100~GHz) and CryoCache model in the cryogenic semiconductor clock domain (4~GHz).
    \item \textit{SuperCore and SuperCache (SuperAll)} - CryoCore and CryoCache models in the superconductor clock domain (100~GHz).
    \item \textit{In-Order CryoCore and CryoCache (In-Order CryoAll)} - In-order CryoCore model and CryoCache model in the cryogenic semiconductor clock domain (4~GHz).
    \item \textit{In-Order SuperCore and CryoCache (In-Order SuperCryo)} - In-order CryoCore model in the superconductor clock domain (100~GHz) and CryoCache model in the cryogenic semiconductor clock domain (4~GHz).
    \item \textit{In-Order SuperCore and SuperCache (In-Order SuperAll)} - In-order CryoCore and CryoCache models in the superconductor clock domain (100~GHz).

    % \item \texttt{Experiment 1} - Varying only the CryoCore clock frequency from 4 GHz to 100 GHz to simulate a superconducting environment. The L1, L2, and L3 caches are kept at 4 GHz to simulate a cryogenic environment.
    % \item \texttt{Experiment 2} - Varying the CryoCore and L1 Cache clock frequencies from 4 GHz to 100 GHz to simulate a superconducting environment. The L2 and L3 caches are kept at 4 GHz to simulate a cryogenic environment.
    % \item \texttt{Experiment 3} - Varying the CryoCore, L1 Cache, and L2 Cache clock frequencies from 4 GHz to 100 GHz to simulate a superconducting environment. The L3 cache is kept at 4 GHz to simulate a cryogenic environment.
    % \item \texttt{Experiment 4} - Varying the CryoCore, L1 Cache, L2 Cache, and L3 Cache clock frequencies from 4 GHz to 100 GHz to simulate a superconducting environment.
\end{itemize}

\section{Results}

% We present the results of our experiments in this section.

Our experimental setup allowed us to answer the following research questions:

% \note{You can use the ``enumerate'' package to reduce the indentation for this enumeration.}
\begin{tcolorbox}[colback=white, colframe=black, sharp corners, title=Research Questions]
    \begin{enumerate}[leftmargin=*]
        \item How does the performance of workloads vary with increasing clock frequency when subsystems are placed in a superconductor environment?
        \item How is the performance of workloads affected by whether the core is out-of-order or in-order?
        \item What new constraints must subsystems meet to leverage the improved performance resulting from increased clock frequency in a superconductor environment?
    \end{enumerate}
\end{tcolorbox}

\subsection{Workload Performance with Increasing Clock Frequency in Superconductor Environments}

\begin{figure}[h]
    \centering
    \begin{subfigure}{\linewidth}
        \centering
        \includegraphics[width=\linewidth]{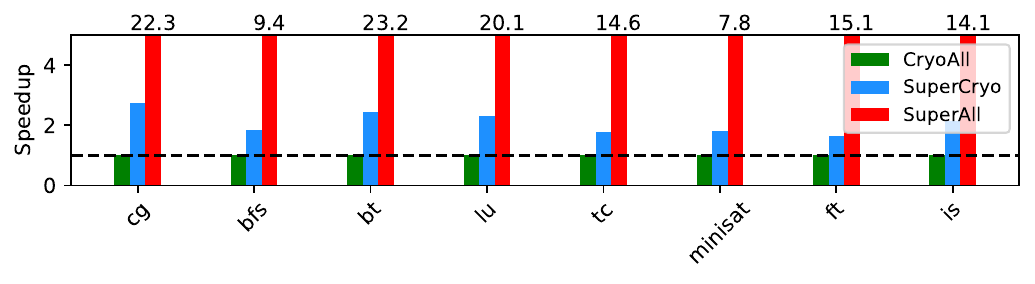}
        \caption{Speedup of small workloads with respect to the out-of-order CryoCore and CryoCache configuration, in different out-of-order setups.}
        \label{fig:outoforder_small}
    \end{subfigure}
    \begin{subfigure}{\linewidth}
        \centering
        \includegraphics[width=\linewidth]{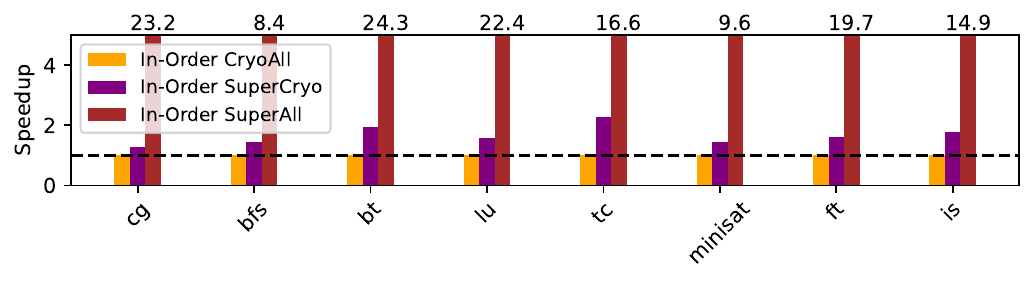}
        \caption{Speedup of small workloads with respect to the in-order CryoCore and CryoCache configuration, in different in-order setups.}
        \label{fig:inorder_small}
    \end{subfigure}
    \begin{subfigure}{\linewidth}
        \centering
        \includegraphics[width=\linewidth]{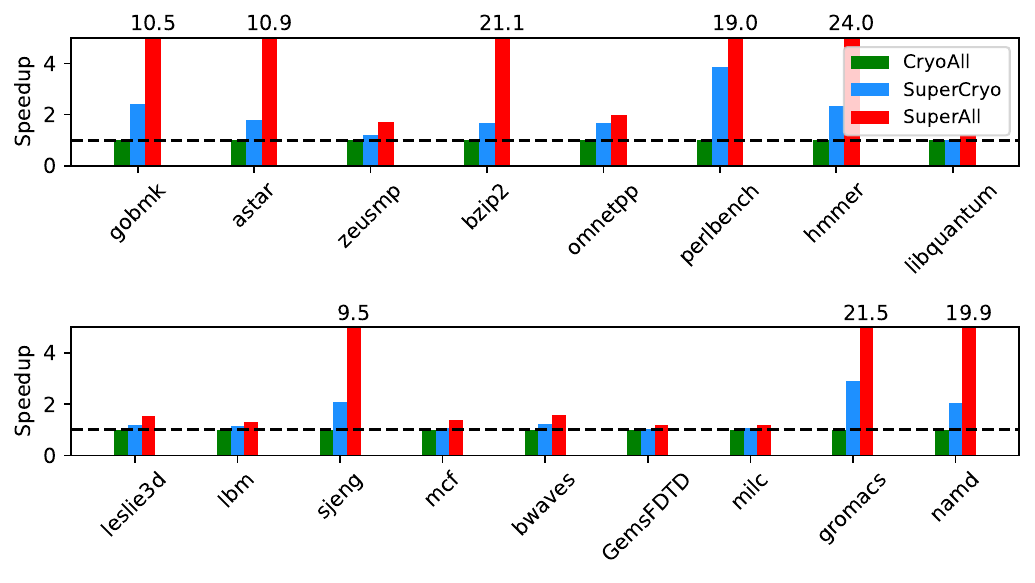}
        \caption{Speedup of large workloads with respect to the out-of-order CryoCore and CryoCache configuration, in different out-of-order setups.}
        \label{fig:outoforder_large}
    \end{subfigure}
    \begin{subfigure}{\linewidth}
        \centering
        \includegraphics[width=\linewidth]{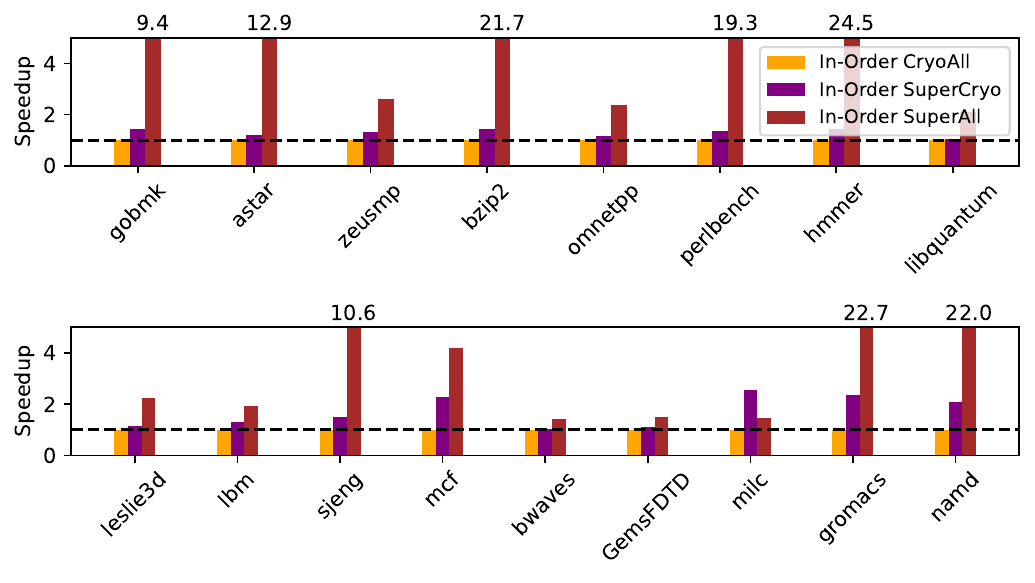}
        \caption{Speedup of large workloads with respect to the in-order CryoCore and CryoCache configuration, in different in-order setups.}
        \label{fig:inorder_large}
    \end{subfigure}
    \caption{Speedup of workloads with respect to their respective core architectures.}
    \label{fig:speedup_arch}
\end{figure}

Figure~\ref{fig:outoforder_small} and~\ref{fig:inorder_small} show the speedup of the small workloads with respect to the out-of-order and in-order cores.
Specifically, it compares the speedups achieved by placing the core and/or cache in a superconductor environment, for both out-of-order and in-order cores.

At a high level, the takeaways from these results are that placing both core and cache in a superconductor environment significantly boosts performance, and that speedups of in-order superconductor components over their cryogenic counterparts are generally higher than the speedups of out-of-order superconductor components over their cryogenic counterparts.
By characterizing the workloads that benefit the most from superconductor components, we also provide insights into the design trade-offs associated with these novel technologies.

For out-of-order cores, Figure~\ref{fig:outoforder_small} shows the speedup of the small workloads in various out-of-order core configurations, with respect to the out-of-order CryoCore and CryoCache configuration.
The bars represent different configurations for different small workloads.
If only the core is placed in the superconductor environment, the performance improvement is not substantial, with the maximum speedup for small workloads being 2.7$\times$ for the \texttt{cg} workload over the baseline for an out-of-order core.
Overall, the speedups for the small workloads range from 1.8$\times$ to 2.7$\times$ for the out-of-order SuperCore and CryoCache configuration, compared to the out-of-order CryoCore and CryoCache configuration.
If both the core and the cache are placed in the superconductor environment, the performance improvement is more substantial for the out-of-order core.
The maximum speedup for the small workloads for an out-of-order SuperCore and SuperCache configuration is 23.2$\times$ for the \texttt{bt} workload over the baseline out-of-order CryoCore and CryoCache configuration.
Some workloads like \texttt{lu} and \texttt{cg} also receive speedups of more than 20$\times$ for the out-of-order SuperCore and SuperCache configuration, compared to the out-of-order CryoCore and CryoCache configuration.
However, some workloads like \texttt{bfs} and \texttt{minisat} do not receive substantial speedups, with the speedup being around 9.4$\times$ and 7.8$\times$ for the out-of-order SuperCore and SuperCache configuration, compared to the out-of-order CryoCore and CryoCache configuration, respectively.
Overall, the speedups for the small workloads range from 7.8$\times$ to 23.2$\times$ for the out-of-order SuperCore and SuperCache configuration, compared to the out-of-order CryoCore and CryoCache configuration.

Figure~\ref{fig:outoforder_large} shows the speedup of the large workloads in various out-of-order core configurations, with respect to the out-of-order CryoCore and CryoCache configuration.
The bars represent different configurations for different SPEC2006 workloads.
Similar to the small workloads, if only the core is placed in the superconductor environment, the performance improvement is not substantial, with the maximum speedup being 3.9$\times$ for the \texttt{400.perlbench} workload over the baseline for an out-of-order core.
Overall, the speedups for the large workloads range from 1.01$\times$ to 3.9$\times$, barring \texttt{456.hmmer}, for the out-of-order SuperCore and CryoCache configuration, compared to the out-of-order CryoCore and CryoCache configuration.
If both the core and the cache are placed in the superconductor environment, the performance improvement is more substantial for the out-of-order core.
The maximum speedup for the large workloads for an out-of-order SuperCore and SuperCache configuration is 24.0$\times$ for the \texttt{456.hmmer} workload over the baseline out-of-order CryoCore and CryoCache configuration.
Some workloads like \texttt{435.gromacs} and \texttt{401.bzip2} also receive speedups of more than 20$\times$ for the out-of-order SuperCore and SuperCache configuration, compared to the out-of-order CryoCore and CryoCache configuration.
However, most workloads do not receive substantial speedups, with the speedup being around 1.2$\times$ and 1.7$\times$ for the out-of-order SuperCore and SuperCache configuration, compared to the out-of-order CryoCore and CryoCache configuration, respectively.
Overall, the speedups for the large workloads range from 1.18$\times$ to 24.0$\times$ for the out-of-order SuperCore and SuperCache configuration, compared to the out-of-order CryoCore and CryoCache configuration.

For in-order cores, Figure~\ref{fig:inorder_small} shows the speedup of the workloads in various in-order core configurations, with respect to the in-order CryoCore and CryoCache configuration.
The bars represent different configurations for different small workloads.
If only the core is placed in the superconductor environment, the performance improvement is not substantial, with the maximum speedup for the small workloads being 2.3$\times$ for the \texttt{tc} workload over the baseline for an in-order core.
Overall, the speedups for the small workloads range from 1.4$\times$ to 2.3$\times$ for the in-order SuperCore and CryoCache configuration, compared to the in-order CryoCore and CryoCache configuration.
If both the core and the cache are placed in the superconductor environment, the performance improvement is more substantial for both the in-order core.
The maximum speedup for an in-order SuperCore and SuperCache configuration is 24.3$\times$ for the \texttt{bt} workload over the baseline in the in-order SuperCore and CryoCache configuration.
Some workloads like \texttt{lu} and \texttt{cg} also receive speedups of more than 20$\times$ for the in-order SuperCore and SuperCache configuration, compared to the in-order CryoCore and CryoCache configuration.
However, some workloads like \texttt{bfs} and \texttt{minisat} do not receive substantial speedups, with the speedup being around 8.4$\times$ and 9.6$\times$ for the in-order SuperCore and SuperCache configuration, compared to the in-order CryoCore and CryoCache configuration, respectively.
Overall, the speedups for the small workloads range from 8.4$\times$ to 24.3$\times$ for the in-order SuperCore and SuperCache configuration, compared to the in-order CryoCore and CryoCache configuration.

Figure~\ref{fig:inorder_large} shows the speedup of the large workloads in various in-order core configurations, with respect to the in-order CryoCore and CryoCache configuration.
The bars represent different configurations for different SPEC2006 workloads.
Similar to the small workloads, if only the core is placed in the superconductor environment, the performance improvement is not substantial, with the maximum speedup being 2.6$\times$ for the \texttt{433.milc} workload over the baseline for an in-order core.
Overall, the speedups for the large workloads range from 1.1$\times$ to 2.6$\times$ for the in-order SuperCore and CryoCache configuration, compared to the in-order CryoCore and CryoCache configuration.
If both the core and the cache are placed in the superconductor environment, the performance improvement is more substantial for both the in-order core.
The maximum speedup for the large workloads for an in-order SuperCore and SuperCache configuration is 24.5$\times$ for the \texttt{456.hmmer} workload over the baseline in the in-order SuperCore and CryoCache configuration.
Some workloads like \texttt{435.gromacs}, \texttt{444.namd} and \texttt{401.bzip2} also receive speedups of more than 20$\times$ for the in-order SuperCore and SuperCache configuration, compared to the in-order CryoCore and CryoCache configuration.
However, most workloads do not receive substantial speedups, with the speedup being around 1.4$\times$ and 2.6$\times$ for the in-order SuperCore and SuperCache configuration, compared to the in-order CryoCore and CryoCache configuration, respectively.
Overall, the speedups for the large workloads range from 1.4$\times$ to 24.5$\times$ for the in-order SuperCore and SuperCache configuration, compared to the in-order CryoCore and CryoCache configuration.

Therefore, the speedups vary significantly across different workloads, with similar workloads achieving similar speedups across different core architectures.
While the speedups are consistent for smaller workloads, the variation in speedup is much wider when running real-world workloads.
This variation suggests that even if we could build a conventional architecture in a superconductor environment, it is unlikely to benefit general-purpose workloads uniformly.
Therefore, we need to either develop a new architecture specifically optimized for superconductor environments or apply superconductor technology to specialized accelerators where there is a small amount of memory traffic instead of general-purpose compute, which could leverage the low latency and high throughput benefits of superconductors more effectively.
% \note{another take away is that while the speedups are consistent for the smaller workloads, when running real-world workloads the speedup varies widely. Thus, it seems like even if we could build a conventional architecture in superconducting, it is unlikely to benefit general-purpose workloads. This conclusion implies that we need to either develop a new architecture or apply superconducting to accelerators where there will be a small amount of memory traffic instead of general-purpose compute.}

\begin{figure}[h]
    \centering
    \begin{subfigure}{\linewidth}
        \centering
        \includegraphics[width=\linewidth]{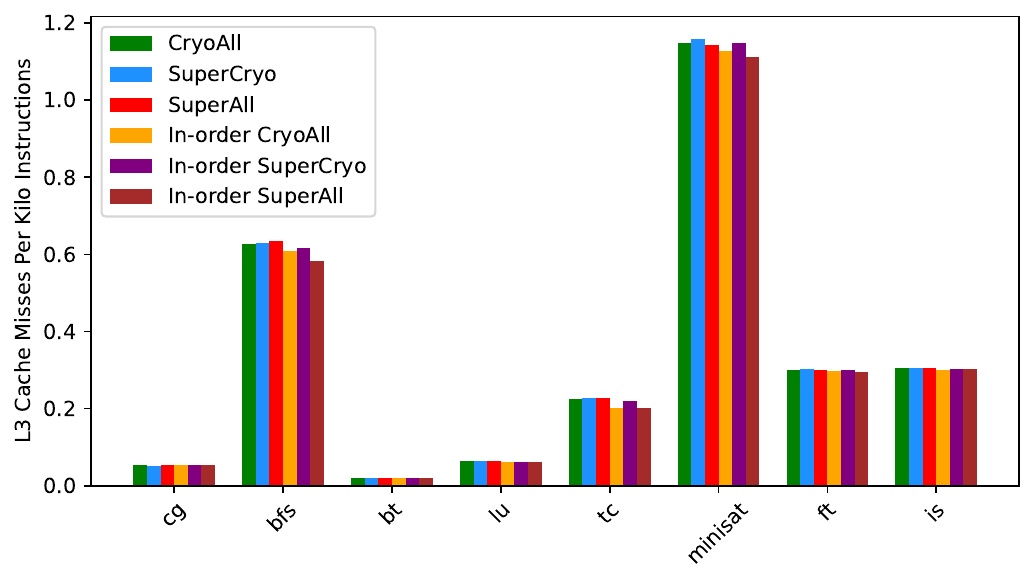}
        \caption{L3 cache misses per kilo instructions (MPKI) for the small workloads.}
        \label{fig:l3misses_small}
    \end{subfigure}
    \begin{subfigure}{\linewidth}
        \centering
        \includegraphics[width=\linewidth]{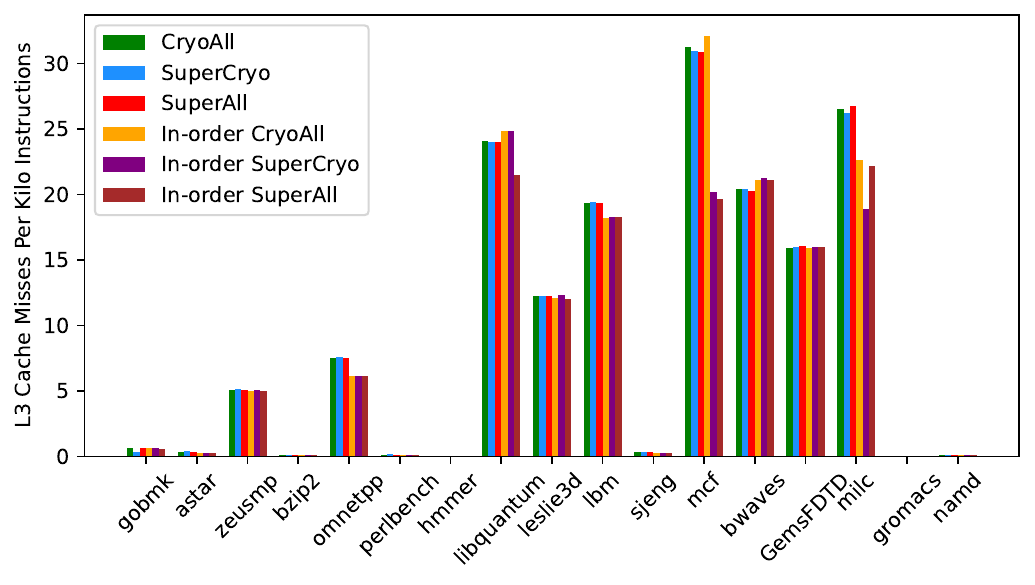}
        \caption{L3 cache misses per kilo instructions (MPKI) for the large workloads.}
        \label{fig:l3misses_large}
    \end{subfigure}
    \caption{L3 cache misses per kilo instructions (MPKI) for the small and large workloads.}
    \label{fig:l3misses}
\end{figure}

Figure~\ref{fig:l3misses_small} shows the L3 cache misses normalized to the number of instructions for the small workloads (L3 cache MPKI).
The bars represent different configurations for different small workloads.
Figure~\ref{fig:l3misses_large} shows the L3 cache misses normalized to the number of instructions for the large workloads (L3 cache MPKI).
The bars represent different configurations for different SPEC2006 workloads.
The L3 cache misses are normalized to the number of instructions to account for the different instruction counts of the workloads.
We notice that the highest L3 cache MPKI for the small workloads are for workloads like \texttt{bfs} and \texttt{minisat}, which also have the lowest speedups.
Similarly, the highest L3 cache MPKI for the large workloads are for workloads like \texttt{429.mcf}, \texttt{434.zeusmp}, \texttt{433.milc}, \texttt{410.bwaves} and \texttt{462.libquantum}, which also have the lowest speedups.
Therefore, the L3 cache misses, and by extension the memory accesses, are a bottleneck for these workloads.
On the other hand, small workloads like \texttt{lu}, \texttt{cg} and \texttt{bt}, and large workloads like \texttt{445.gobmk}, \texttt{400.perlbench}, \texttt{456.hmmer}, \texttt{444.namd}, and \texttt{473.astar} have lower normalized L3 cache misses and higher speedups, indicating that these workloads are not bottlenecked by memory accesses, and therefore, are able to achieve higher speedups.

Therefore, we conclude that the workloads which do not show substantial speedups are bottlenecked by memory accesses.
This is because the memory operates at room temperature, and the core has to wait for the memory to respond to its requests, implying that the workload would not be able to achieve the speedup that the faster core could achieve.
This is a potential limitation of the system, as the memory system would need to be redesigned to keep up with the faster core.

\begin{tcolorbox}[colback=white, colframe=black, sharp corners, title=Takeaways]
    \begin{itemize}[leftmargin=*]
        \item The performance of workloads improved by placing components in the superconductor environment.
        The performance improvement was more substantial when both the core and the cache were in the superconductor environment.
        \item In-order cores were able to achieve a higher speedup than out-of-order cores when both the core and the cache were placed in the superconductor environment.
        \item The workloads that did not receive substantial speedups were bottlenecked by memory accesses.
        Therefore, memory systems would need to be redesigned to keep up with the faster core.
        \item The speedups vary significantly across different workloads, with similar workloads achieving similar speedups across different core architectures.
        While the speedups are consistent for smaller workloads, the variation in speedup is much wider when running real-world workloads.
        Therefore, we need to either develop a new architecture specifically optimized for superconductor environments or apply superconductor technology to specialized accelerators where there is a small amount of memory traffic instead of general-purpose compute.
    \end{itemize}
\end{tcolorbox}

\subsection{Effect of Core Type (Out-of-Order vs. In-Order) on Workload Performance}

\begin{figure}[h]
    \centering
    \begin{subfigure}{\linewidth}
        \centering
        \includegraphics[width=\linewidth]{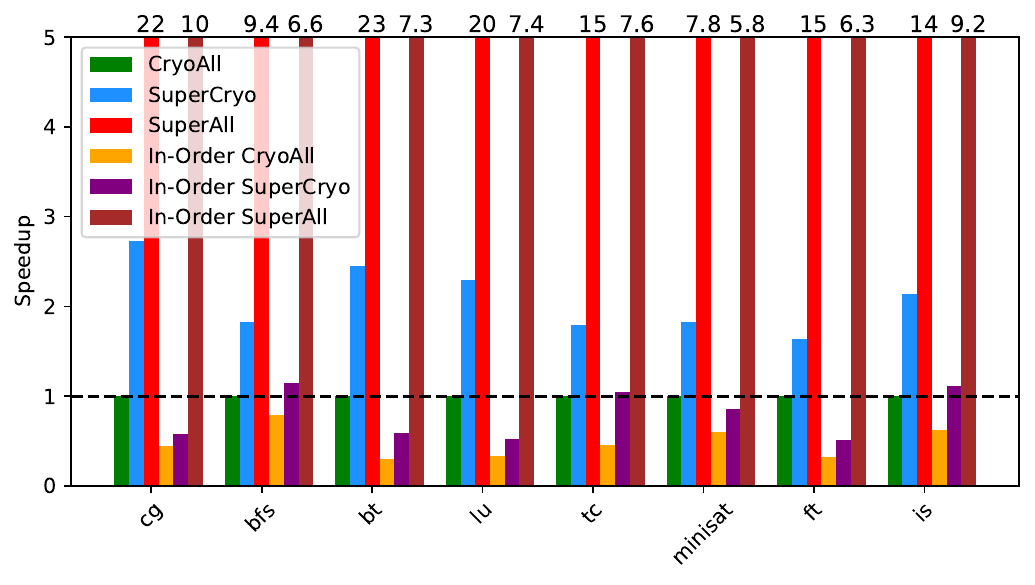}
        \caption{Speedup of small workloads with respect to the out-of-order CryoCore and CryoCache configuration, in different setups.}
        \label{fig:speedup_small}
    \end{subfigure}
    \begin{subfigure}{\linewidth}
        \centering
        \includegraphics[width=\linewidth]{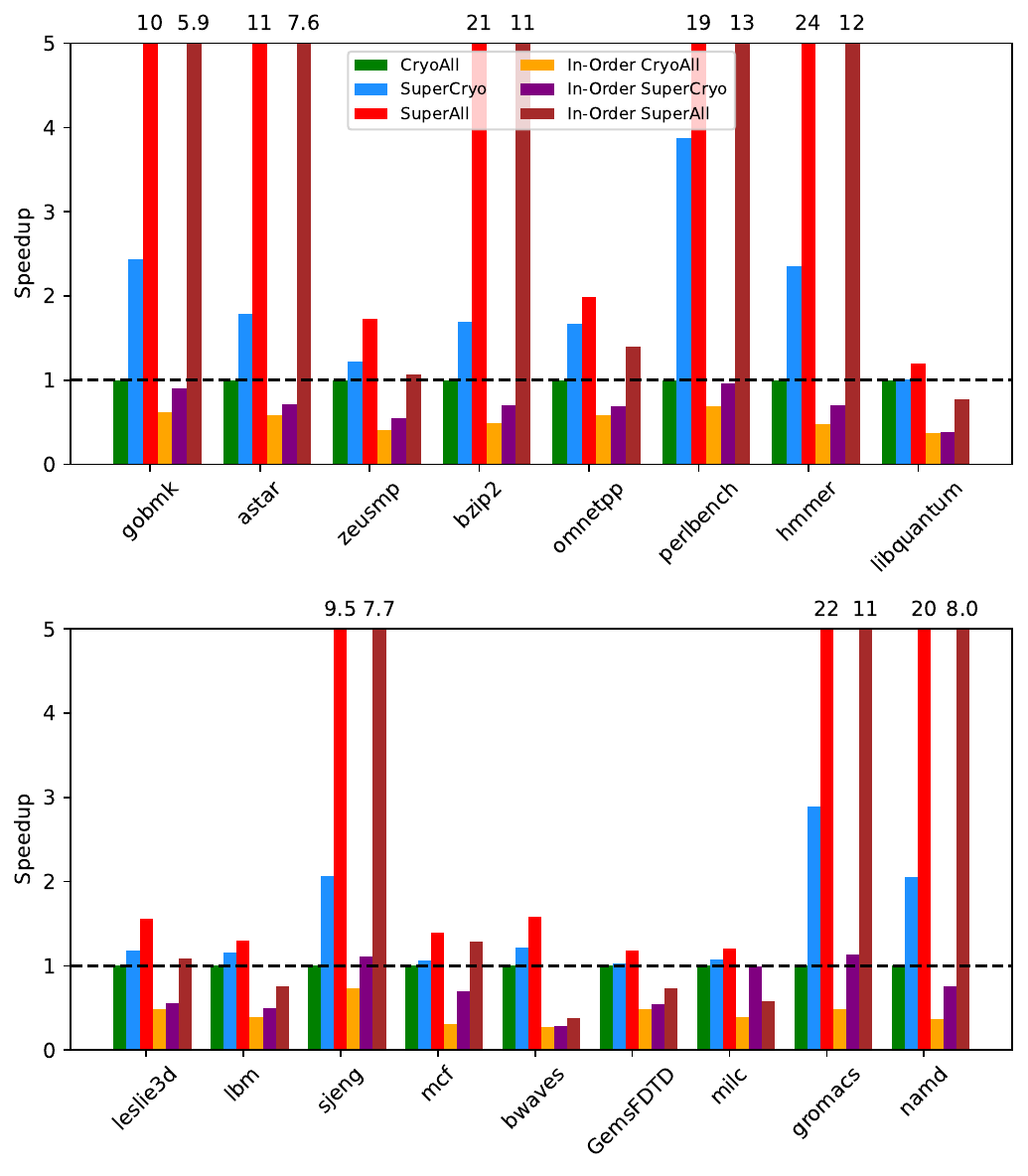}
        \caption{Speedup of large workloads with respect to the out-of-order CryoCore and CryoCache configuration, in different setups.}
        \label{fig:speedup_large}
    \end{subfigure}
    \caption{Speedup of workloads with respect to the out-of-order CryoCore and CryoCache configuration, in different setups.}
    \label{fig:speedup}
\end{figure}

We compared the speedups of various configurations of out-of-order and in-order superconductor and cryogenic semiconductor cores and caches, with respect to the out-of-order CryoCore and CryoCache configuration.
At a high level, we observed that the speedups are more substantial for out-of-order configurations compared to in-order configurations.

Figure~\ref{fig:speedup_small} shows the speedup of the workloads with respect to the out-of-order CryoCore and CryoCache configuration, in different setups.
The bars represent different configurations for different small workloads. %, including \texttt{bfs}, \texttt{tc}, \texttt{minisat}, \texttt{is}, \texttt{lu}, \texttt{cg}, \texttt{bt}, and \texttt{ft}.
Figure~\ref{fig:speedup_large} shows the speedup of the workloads with respect to the out-of-order CryoCore and CryoCache configuration, in different setups.
The bars represent different configurations for different SPEC2006 workloads.
Comparing the speedups of both the out-of-order and in-order cores with respect to the out-of-order CryoCore and CryoCache configuration, we see that the speedups are more substantial for the out-of-order cores, for both the small and large workloads.
The out-of-order SuperCore and CryoCache configuration performs better than or equal to both the in-order CryoCore and CryoCache configuration and the in-order SuperCore and CryoCache configuration, for both the small and large workloads.

From these figures, we can take away that for small workloads, placing components in a superconductor environment improves performance for both out-of-order and in-order cores.
The performance improvement is more substantial when both the core and cache are in the superconductor environment.
Interestingly, in-order cores achieve higher speedups compared to out-of-order cores when both components are superconductor.
This could be because in-order cores are simpler and have less instruction-level parallelism, which could benefit from the higher clock frequency.

The out-of-order SuperCore and CryoCache configuration performs better than or equal to both the in-order CryoCore and CryoCache configuration and the in-order SuperCore and SuperCache configuration, for both the small and large workloads.
Out-of-order cores can schedule instructions dynamically, which could benefit more from the higher clock frequency compared to in-order cores, which have a fixed schedule of instructions.
Thus, even when all components are not superconductor, the out-of-order cores can still achieve significant performance benefits compared to in-order cores.

\begin{tcolorbox}[colback=white, colframe=black, sharp corners, title=Takeaways]
    \begin{itemize}[leftmargin=*]
        \item The speedups are more substantial for out-of-order configurations compared to in-order configurations.
        \item The out-of-order SuperCore and CryoCache configuration performs better than or equal to both the in-order CryoCore and CryoCache configuration and the in-order SuperCore and CryoCache configuration, for both the small and large workloads.
    \end{itemize}
\end{tcolorbox}

\subsection{New Constraints for Leveraging Improved Performance from Increased Clock Frequency}

\begin{figure}[h]
    \centering
    \begin{subfigure}{\linewidth}
        \centering
        \includegraphics[width=\linewidth]{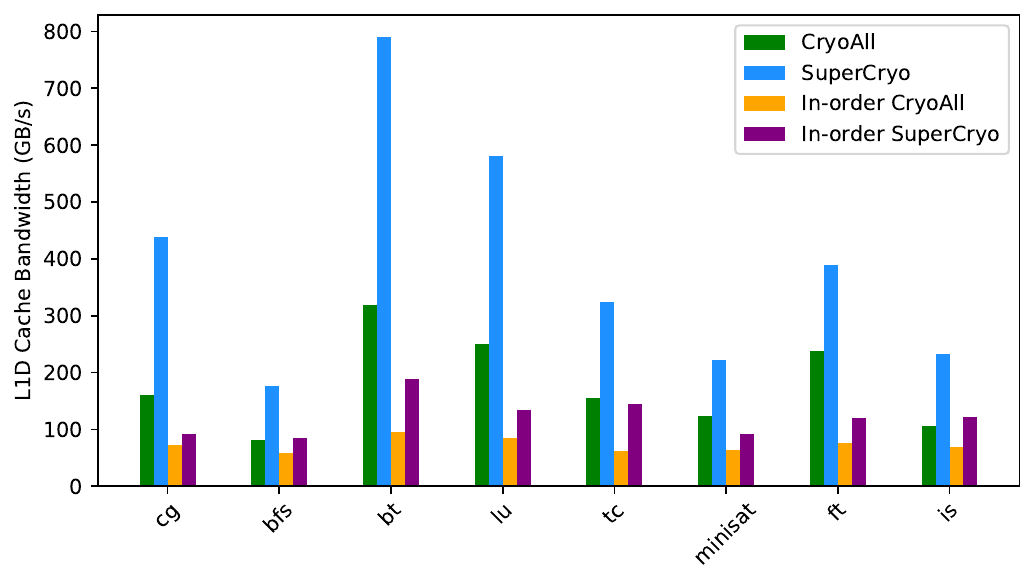}
        \caption{L1D cache bandwidth for the small workloads.}
        \label{fig:l1d_bandwidth_small}
    \end{subfigure}
    \begin{subfigure}{\linewidth}
        \centering
        \includegraphics[width=\linewidth]{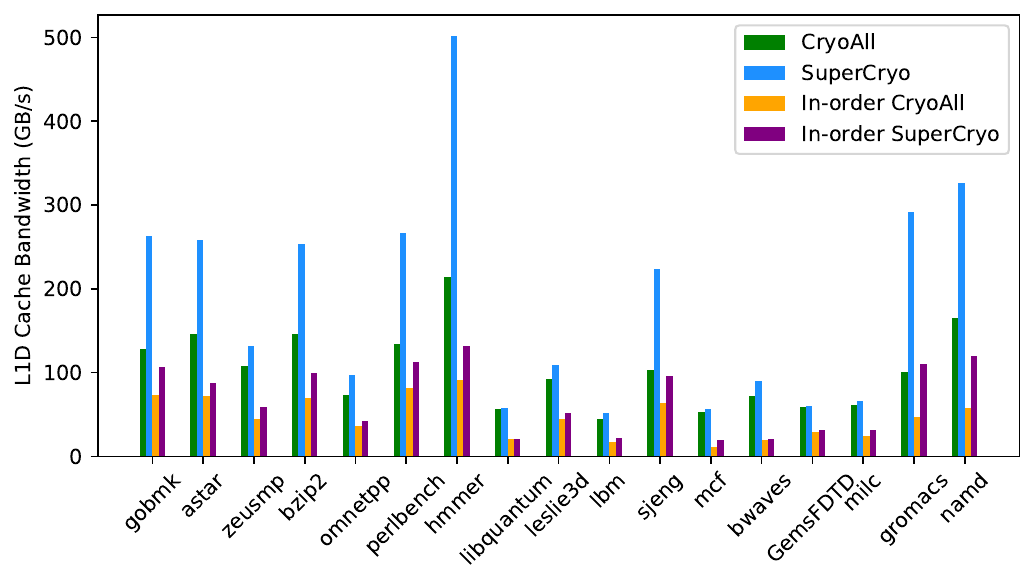}
        \caption{L1D cache bandwidth for the large workloads.}
        \label{fig:l1d_bandwidth_large}
    \end{subfigure}
    \caption{L1D cache bandwidth for the small and large workloads.}
    \label{fig:l1d_bandwidth}
\end{figure}

In order to leverage the improved performance resulting from increased clock frequency in a superconductor environment, subsystems must meet new constraints.
One such constraint is the cache bandwidth required to keep up with the faster core.
We evaluated the cache bandwidths required for the L1I cache, L1D cache, L2 cache, and L3 cache for the small and large workloads.

Figure~\ref{fig:l1d_bandwidth_small} shows the L1D cache bandwidth for the small workloads.
The bars represent different configurations for different small workloads.
Figure~\ref{fig:l1d_bandwidth_large} shows the L1D cache bandwidth for the large workloads.
The bars represent different configurations for different SPEC2006 workloads.
The configurations in this figure are the CryoCore and CryoCache configuration, the SuperCore and CryoCache configuration, and their in-order variants.
We chose these because they are more ``realistic'' than having a configuration with a superconductor cache, since memory systems for superconductor are not as scalable yet~\cite{alam2023cryogenic}.
We notice that the L1D cache bandwidth is higher for the SuperCore and CryoCache configuration compared to the CryoCore and CryoCache configuration.
For the small workloads, the maximum bandwidth required for the out-of-order architecture is 800 GB/s for the \texttt{bt} workload in the SuperCore and CryoCache configuration, and for in-order architecture, it is 200 GB/s for the \texttt{bt} workload in the in-order SuperCore and CryoCache configuration.
For the large workloads, the maximum bandwidth required for the out-of-order architecture is 500 GB/s for the \texttt{456.hmmer} workload in the SuperCore and CryoCache configuration, and for in-order architecture, it is 130 GB/s for the \texttt{456.hmmer} workload in the in-order SuperCore and CryoCache configuration.

We observe higher L1D cache bandwidth for the SuperCore and CryoCache configuration compared to the CryoCore and CryoCache configuration, for both the small and large workloads, for both the out-of-order and in-order cores.
The higher clock frequency of the core in the SuperCore configuration results in more requests to the L1D cache, which increases the number of accesses per unit time, and hence the bandwidth.
The L1I caches, L2 caches, and L3 cache bandwidths are also higher for the SuperCore and CryoCache configuration compared to the CryoCore and CryoCache configuration, for both the small and large workloads, for a similar reason.
For the small workloads, the maximum bandwidth required for the L1I cache is 190 GB/s for the \texttt{tc} workload in the SuperCore and CryoCache configuration; for the L2 cache, it is 390 GB/s for the \texttt{ft} workload in the SuperCore and CryoCache configuration; and for the L3 cache, it is 190 GB/s for the \texttt{cg} workload in the SuperCore and CryoCache configuration.
For the large workloads, the maximum bandwidth required for the L1I cache is 130 GB/s for the \texttt{400.perlbench} workload in the SuperCore and CryoCache configuration; for the L2 cache, it is 120 GB/s for the \texttt{435.gromacs} workload in the SuperCore and CryoCache configuration; and for the L3 cache, it is 70 GB/s for the \texttt{470.lbm} workload in the SuperCore and CryoCache configuration.
% However, the highest bandwidth requirements are for the L1D cache, which is consistent with the fact that the L1D cache is the first level of cache that the core accesses, and hence the most frequently accessed cache.
The workloads with the highest speedups, both in the small and large workloads, have the highest L1D cache bandwidths.
If these bandwidths are not met, the performance improvement would not be actualized.
Therefore, these high bandwidth requirements are a potential limitation of the system, and require redesigning the caches to keep up with the faster core.

\begin{tcolorbox}[colback=white, colframe=black, sharp corners, title=Takeaways]
    \begin{itemize}[leftmargin=*]
        \item The workloads with the highest speedups have the highest L1D cache bandwidths.
        \item The overall required cache bandwidths are in the range of 130 GB/s to 800 GB/s for the small workloads and 70 GB/s to 500 GB/s for the large workloads.
        If these bandwidths are not met, the performance improvement would not be actualized.
    \end{itemize}
    Therefore, these high bandwidth requirements are a potential limitation of the system, and require redesigning the caches to keep up with the faster core.
\end{tcolorbox}

\section{Limitations}

We only observed the effects of high-level implications of superconductor components and cryogenic semiconductor components on the performance of workloads, i.e., the higher clock frequency of the components in these environments.
We did not consider superconductor components having a greater pipeline depth~\cite{zha2023superbp} nor temporal logic in superconductor circuits~\cite{tzimpragos2020computational, tzimpragos2021temporal}, which has an impact on the performance of the workloads.
We also did not model the SERDES (serializer-deserializer) circuits that would be required to interface the superconductor components with the room-temperature components, which would have an impact on the performance of the workloads.
Instead, we assumed that the interconnect is unchanged from CMOS.

The SuperCore and SuperCache configuration is not feasible in practice.
Our paper is a thought experiment to see the potential speedups that could be achieved if an ideal 100~GHz superconductor core and cache were used.
In practice, the memory systems for superconductor are not as scalable yet~\cite{alam2023cryogenic}, and the bandwidth requirements limit the performance of the workloads.

\section{Related Work}

Cryogenic semiconductor computing has been a topic of interest for researchers for a long time.
Our work is based on the work of Byun et al.~\cite{cryocore} and Min et al.~\cite{min2020cryocache}, who have proposed a cryogenic core and a cryogenic cache, respectively.
Lee et al. have also designed a cryogenic memory~\cite{lee2019cryogenic}, which could be used in addition to the cryogenic core and cache to create a complete cryogenic semiconductor computing system.
In this work, we have adapted their designs to gem5 and conducted simulations to evaluate the performance of workloads in cryogenic semiconductor computing environments.

There has been some work on superconductor computing cores as well.
Ando et al. have proposed a design of an 8-bit Microprocessor based on Rapid Single Flux Quantum (RSFQ) technology, called ``CORE e4''~\cite{coree4}.
Yamanashi et al. have proposed a design of a pipelined 8-bit microprocessor based on Single Flux Quantum (SFQ) technology, called ``CORE1$\beta$''~\cite{core1beta}.
FLUX Chip by Dorojevets et al. is a chip that uses RSFQ technology~\cite{fluxchip}.
It is a 16-bit microprocessor that can be clocked at 20 GHz.
In this work, we just look at the high-level implication of superconductor (i.e., a clock frequency of 100 GHz) core and cache on the performance of workloads in a superconductor environment in gem5.

There has also been some work on superconductor components, part of a larger system, that may be in a room-temperature or cryogenic environment.
Herr et al. have proposed a design of a 4-bit RSFQ multiplier-accumulator~\cite{herr1997design}.
They ran tests on its speed and power consumption, and found that was clocked at 11 GHz and consumed 0.2 mW.
Zha et al. have proposed a superconductor perceptron-based branch predictor~\cite{zha2023superbp}.
Nagaoka et al. have proposed a design of a bit-parallel multiplier based on RSFQ technology, which can clock at 52 GHz~\cite{nagaoka2021demonstration}.
Nagaoka et al. also proposed a design of a gate-leveled SFQ multiplier, which can clock at 48 GHz~\cite{nagaoka201929}.
Obata et al. have proposed an SFQ integer multiplier with a systolic array architecture, clocked at 25 GHz~\cite{obata2006single}.
In this work, we just look at the an entire core or cache in a superconductor environment, and its impact on the performance of workloads.
We do not look at the performance of individual components, like a multiplier or a branch predictor, in a superconductor environment in gem5.

There has also been some work on temporal logic in superconductor circuits.
Tzimpragos et al.~\cite{tzimpragos2020computational, tzimpragos2021temporal} propose that superconductor circuits can compute over temporal relationships between pulse arrivals.
They propose the computational relationships between those pulse arrivals can be formalized through an extension to a temporal predicate logic.
In this work, we do not look at the temporal logic in superconductor circuits, but rather the high-level implications of a superconductor core and cache on the performance of workloads in a superconductor environment in gem5.

\section{Conclusion}

In this paper, we presented our methodology for conducting simulations of workloads in cryogenic semiconductor computing environments using gem5.
We then presented the results of our experiments and discussed their implications for the development of cryogenic semiconductor computing technologies.

Our experiments show that gem5 can be used to simulate workloads in cryogenic semiconductor computing environments.
We used this to show that the performance of the workloads does improve with increasing CryoCore and CryoCache clock frequencies, and characterize the workloads that do not show a substantial improvement in performance.
We also showed that the caches would need to be able to provide a higher bandwidth to the core to achieve the speedups shown in the experiments, and provided the bandwidths required to achieve these speedups.

% Our work provides valuable insights for the design and optimization of future cryogenic computing systems.
% We hope that our work will inspire future research and development in this field.

\section*{Acknowledgments}

We would like to thank the members of the Davis Architecture Research Group (DArchR) for their valuable feedback and support throughout this project.
Particularly, we would like to thank Zhantong Qiu for her help in setting up the experiments related to the SimPoints of the large workloads, Harshil Patel for his help in creating the suite of small workloads used in the experiments, and Mahyar Samani for providing feedback on initial drafts of this paper.
We would also like to thank D. Scott Holmes and the members of the IEEE IRDS CEQIP team for their valuable feedback on an initial version of this paper.
This work was funded by the Department of Energy under grant number DE-SC0024502.

\bibliography{references}

\bibliographystyle{IEEEtran}

\end{document}